\def\be{\begin{equation}}
\def\ee{\end{equation}}
\def\bea{\begin{eqnarray}}
\def\eea{\end{eqnarray}}
\documentclass{appolb}
\usepackage{epsfig}

\begin{document}
\title{Squeezed particle-antiparticle correlations%
\thanks{Presented at the IV Workshop on Particle Correlations and Femtoscopy (WPCF 2008), Krakow, Poland, September 11-14, 2008. } 
}
\author{Sandra S. Padula, Danuce M. Dudek,
\address{IFT-UNESP, Rua Pamplona, 145, 01405-900 S\~ao Paulo, SP, Brazil
}
\and
 Ot\'avio Socolowski Jr.
\address{IMEF - FURG - C. P. 474, 96201-900, Rio Grande, RS, Brazil}
}
\maketitle
\begin{abstract} A novel type of correlation involving particle-anti\-particle pairs was 
found out in the 1990's. Currently known as squeezed or Back-to-Back Correlations (BBC), they should be present 
if the hadronic masses are modified in the hot and dense medium formed in high energy heavy ion collisions. 
Although well-established theoretically, such hadronic correlations have not yet been observed experimentally.  In this phenomenological study we suggest a promising way to search for the BBC signal,  by looking into the squeezed correlation function of $\phi \phi$ and $K^+ K^-$ pairs at RHIC energies, as function of the pair average momentum, $\mathbf{K_{12}}=(\mathbf{k_1}+\mathbf{k_2})/2$. The effects of in-medium mass-shift on the  identical particle correlations (Hanbury-Brown \& Twiss effect) are also discussed.
\end{abstract}
\PACS{25.75.Gz, 25.75.-q, 21.65.Q, 21.65.Jk}
 
 \section{Introduction}
The hadronic particle-antiparticle correlation was already pointed out in the beginning of the nineties. However, the final formulation of these hadronic  squeezed or back-to-back correlations was proposed only at the end of that decade\cite{acg99}, predicting that such correlations were expected if the masses of the mesons were modified in the hot and dense medium formed in 
high energy nucleus-nucleus collisions. Soon after that, it was shown that analogous correlations would exist in the case of baryons as well. An interesting theoretical finding was that  both the fermionic (fBBC) and the bosonic (bBBC)  Back-to-Back Correlations were very similar, both being positive and treated by analogous formalisms. 
In what follows, we will focus our discussion to the bosonic case, illustrating the effect by considering 
$\phi \phi$ and $K^+ K^-$ pairs, considered to be produced at RHIC energies\cite{phkpc05}.

Let us discuss the case of $\phi$-mesons first, which are their own antiparticles, and suppose that their masses 
are modified in hot and dense  medium. Naturally, they recover their asymptotic masses after the 
system freezes-out. Therefore,  the joint probability for observing two such particles, i.e., the two-particle distribution, 
 $N_2(\mathbf k_1,\mathbf k_2) = \omega_{\mathbf k_1}  \omega_{\mathbf k_2} \, \langle
a^\dagger_{\mathbf k_1} a^\dagger_{\mathbf k_2} a_{\mathbf k_2} a_{\mathbf k_1} \rangle$, can be factorized as
$N_2(\mathbf k_1,\mathbf k_2)  =  \omega_{\mathbf k_1} \omega_{\mathbf k_2} \, \Bigl[\langle
a^\dagger_{\mathbf k_1} a_{\mathbf k_1}\rangle \langle a^\dagger_{\mathbf k_2}
a_{\mathbf k_2} \rangle + \langle a^\dagger_{\mathbf k_1} a_{\mathbf k_2}\rangle
\langle a^\dagger_{\mathbf k_2} a_{\mathbf k_1} \rangle + \langle
a^\dagger_{\mathbf k_1} a^\dagger_{\mathbf k_2} \rangle \langle a_{\mathbf k_2}
a_{\mathbf k_1} \rangle\Bigr]$,  after applying a generalization of Wick's theorem for locally
equilibrated systems\cite{gykw,sm1}. 
The first term corresponds to the product of the spectra of the two $\phi$'s, 
$N_1(\mathbf k_i)\!=\!\omega_{\mathbf k_i} \frac{d^3N}{d\mathbf k_i} \!=\!
\omega_{\mathbf k_i}\,
\langle a^\dagger_{\mathbf k_i} a_{\mathbf k_i} \rangle $, 
being $a^\dagger_\mathbf k$ and $a_\mathbf k$ the free-particle creation and
annihilation operators of scalar quanta, and 
$\langle ... \rangle $ means thermal averages. 
The second term contains the identical particle contribution and is represented by the 
square modulus of the chaotic amplitude,
$ G_c({\mathbf k_1},{\mathbf k_2}) = \sqrt{\omega_{\mathbf k_1} \omega_{\mathbf k_2}} \; \langle
a^\dagger_{\mathbf k_1} a_{\mathbf k_2} \rangle$. Together with the first term, it 
gives rise to the femtoscopic or Hanbury-Brown \& Twiss (HBT) effect. 
The third term, the square modulus of the squeezed amplitude, 
$G_s({\mathbf k_1},{\mathbf k_2}) = \sqrt{\omega_{\mathbf k_1} \omega_{\mathbf k_2} } \; \langle
a_{\mathbf k_1} a_{\mathbf k_2} \rangle$, 
is identically zero in the absence of in-medium mass-shift. However, if the particle's mass is modified,  
together with the first term it leads to the squeezing correlation function. 

The annihilation (creation) operator of the asymptotic, observed bosons with momentum 
$k^\mu\!=\!(\omega_k,{\bf k})$, $a$ ($a^\dagger$),  is related to the in-medium annihilation (creation) 
operator $b$ ($b^\dagger$), corresponding to thermalized quasi-particles, by the Bogoliubov-Valatin 
transformation,  
$a_k=c_k b_k + s^*_{-k} b^\dagger_{-k} \; ; \; a^\dagger_k=c^*_k
b^\dagger_k + s_{-k} b_{-k}$, where $c_k=\cosh(f_k)$, $s_k=\sinh(f_k)$. The argument,  
$f_{i,j}(x)=\frac{1}{2}\log\left[\frac{K^{\mu}_{i,j}(x)\, u_\mu
(x)} {K^{*\nu}_{i,j}(x) \, u_\nu(x)}\right]$, is 
the  {\sl squeezing parameter}. In terms of the above amplitudes, the complete  $\phi \phi$ correlation function can be written as
\be 
C_2({\mathbf k_1},{\mathbf k_2}) =1 + 
\frac{|G_c({\mathbf k_1},{\mathbf k_2})|^2}{G_c({\mathbf k_1},{\mathbf k_1})
G_c({\mathbf k_2},{\mathbf k_2})} + \frac{|G_s({\mathbf k_1},{\mathbf k_2})
|^2}{G_c({\mathbf k_1},{\mathbf k_1}) G_c({\mathbf k_2},{\mathbf k_2}) }, \label{fullcorr}
\ee 
where the first two terms correspond to the identical particle (HBT) correlation, whereas the first and 
the last terms represent the correlation function between the particle and its antiparticle, i.e., the squeezed part. 
The in-medium modified mass, $m_*$, is related to the asymptotic mass, $m$, by
$m_*^2(|{\bf k}|)=m^2-\delta M^2(|{\bf k}|)$, here assumed to be a constant mass-shift. 

\section{\bf Results} 
 
The formulation for both bosons and fermions was initially derived for a static, infinite medium \cite{acg99,pchkp01}. More  recently, it was shown\cite{phkpc05}  in the bosonic case that, for finite-size systems expanding with moderate flow,  the squeezed correlations may survive with sizable  strength to be observed experimentally. Similar behavior is expected in the fermionic case. In that analysis, a non-relativistic treatment with flow-independent squeezing parameter was adopted for the sake of simplicity, allowing to obtain analytical results. The detailed discussion is in Ref. \cite{phkpc05}, where the maximum value of $C_s({\mathbf k},-{\mathbf k})$, was studied as a 
function of the modified mass, $m_*$, considering pairs with exact back-to-back momentum, 
${\mathbf k_1}\!\!=\!\!-{\mathbf k_2}\!\!=\!\!{\mathbf k}$ (in the identical particle case, this procedure would be analogous to 
study the behavior of the intercept of the HBT correlation function).  
Although illustrating many points of theoretical interest, this study in terms of the unobserved shifted mass and exactly back-to-back momenta 
was not helpful for motivating the experimental search of the BBC's. A more realistic analysis would involve combinations of the momenta of the individual particles, 
$(\mathbf k_1,\mathbf k_2)$, into the average momentum of the pair, $\mathbf K\!=\!\frac{1}{2}( \mathbf k_1+\mathbf k_2)$. Since the maximum of the BBC effect is reached when ${\mathbf k_1}\!=\!-\!{\mathbf k_2}\!=\!{\mathbf k}$, this would correspond to investigate the squeezed correlation function, $C_s({\mathbf k_1},{\mathbf k_2})=C_s({\mathbf K},{\mathbf q})$, close to $|\mathbf K|\!=\!0$. 

For a hydrodynamical ensemble, both the chaotic and the squeezed amplitudes, $G_c$ and $G_s$, respectively, 
can be written in a special form derived in  \cite{sm1} 
and developed in \cite{acg99,phkpc05}. Therefore, within a non-relativistic treatment with flow-independent squeezing parameter, 
the squeezed amplitude is written as in \cite{phkpc05}, i.e., 
{\small
$G_s(\mathbf{k}_1,\mathbf{k}_2)  =  \frac{E_{_{1,2}}}{(2\pi)^\frac{3}{2}} c_{_0}s_{_0} \Bigl\{ R^3  \exp\Bigl[-\frac{R^2}{2}(\mathbf{k}_1+\mathbf{k}_2)^2\Bigr] +  2 n^*_0  R_*^3  \exp\Bigl[-\frac{(\mathbf{k}_1-\mathbf{k}_2)^2}{8m_* T}\Bigr] 
 \exp \Bigl[-\frac{im\langle u\rangle R(\mathbf{k_1} + \mathbf{k_2})^2}{2m_* T_*}\Bigr]   \exp\Bigl[-
\Bigl(  \frac{1}{8 m_* T_*}  +  \frac{R_*^2}{2} \Bigr) (\mathbf{k_1} + \mathbf{k_2})^2\Bigr] \Bigl\} $},  
and the spectrum, as 
{\small$G_c(\mathbf{k}_i,\mathbf{k}_i)  =  \frac{E_{i,i}}{(2\pi)^\frac{3}{2}}  \Bigl\{|s_{_0}|^2R^3  +  n^*_0 R_*^3 
( |c_{_0}|^2 +  |s_{_0}|^2)  \exp\Bigl(-\frac{\mathbf{k}_i^2}{2m_* T_*}\Bigr)  \Bigr\}$, }
where {\small $R_*=R\sqrt{T/T_*}$ and $T_*=T+\frac{m^2\langle u\rangle^2}{m_*}$} \cite{phkpc05,pscn08}. 
We adopt here $\hbar=c=1$. 

Inserting these expressions into Eq. (\ref{fullcorr}) and considering the region where the HBT correlation is not relevant, we 
obtain the results shown in Figure 1.  Part (a) shows the squeezed correlation as a function of $2 \mathbf {K}=(\mathbf{k_1}+\mathbf{k_2})$, for several values of  $\mathbf {q}=(\mathbf{k_1}-\mathbf{k_2})$. The top plot shows results expected in the case of a instant emission of the $\phi \phi$ correlated pair. If, however, the emission happens in a finite interval, the second term in Eq. (\ref{fullcorr}) is multiplied by a reduction factor, in this case expressed by a Lorentzian ({\small $F(\Delta t)=[1+(\omega_1+\omega_2)^2 \Delta t ^2]^-1$}), i.e., the Fourier transform of an exponential emission.  The result is shown in the plot in the middle of Figure 1(a). We see that this represents a dramatic reduction in the signal, even though its strength is sizable for being observed experimentally. If the system expands with radial flow ($\langle u \rangle=0.5$), the result is shown in the plot at the bottom of Figure 1(a), again considering that the $\phi$'s are emitted during a finite period of time, $\Delta t=2$ fm/c. We see that, in the absence of flow, the squeezed correlation signal grows faster for higher values $|\mathbf q|$ than the corresponding case in the presence of flow. However, this last one is stronger in all the investigated $|\mathbf q|$ region, showing that the presence of radial flow enhances the signal. 
The sensitivity of the squeezed-pair correlation to the size of the region where the mass-shift occurs is shown in Figure1(b) for two values of radii, $R=7$ fm and $R=3$ fm, keeping $|\mathbf{q_{_{12}}}|=2.0$ GeV/c fixed. The differences are reflected in the inverse width of the curves, 
plotted as a function of $2 |\mathbf{K}|$. 
In case of no in-medium mass modification, the squeezed correlation functions would be unity for all values of $2 |\mathbf K|$ in both plots. \begin{figure}[htb]
\begin{center}
\includegraphics[height=.36\textheight]{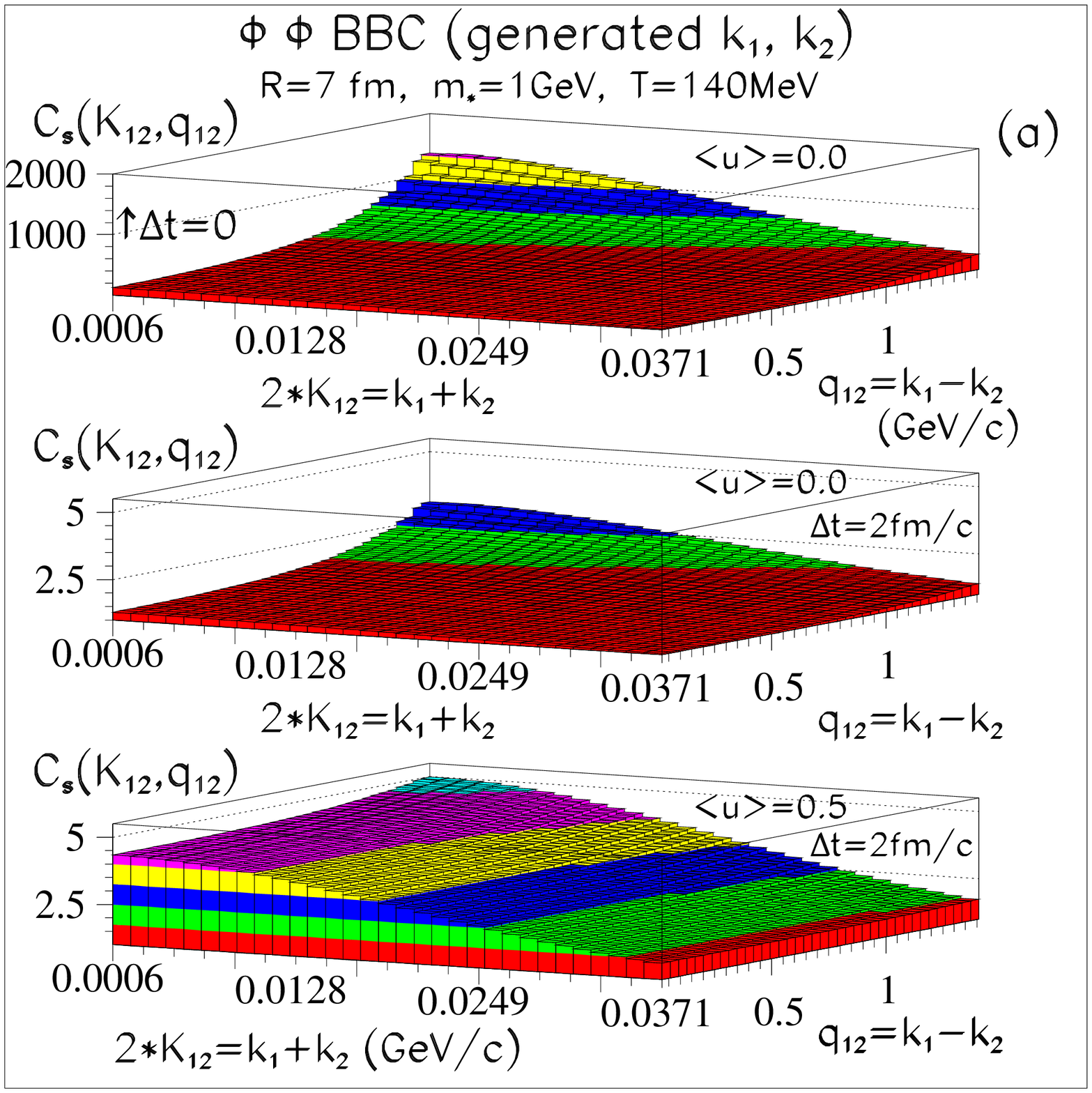}
\includegraphics[height=.41\textheight]{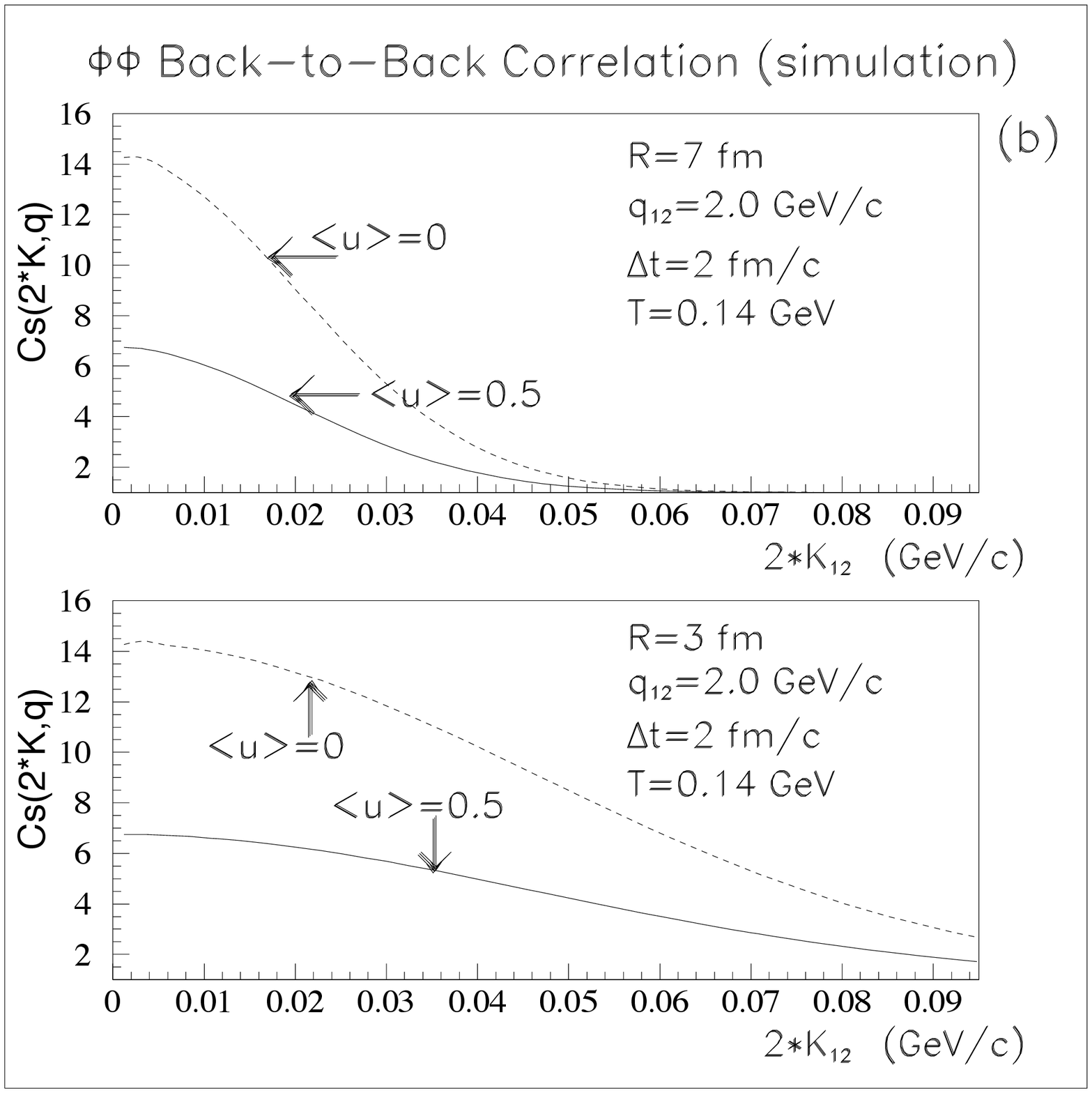}
\caption{ In part (a), the squeezed-pair correlations are shown, illustrating the effects of flow and finite emission times.  Part (b) shows the response of the BBC function to the size of the squeezing region, with $R=7$ fm (top) and $R=3$ fm (bottom).}
\end{center}
\end{figure}

\begin{figure}[htb]
\begin{center}
 \includegraphics[height=.36\textheight]{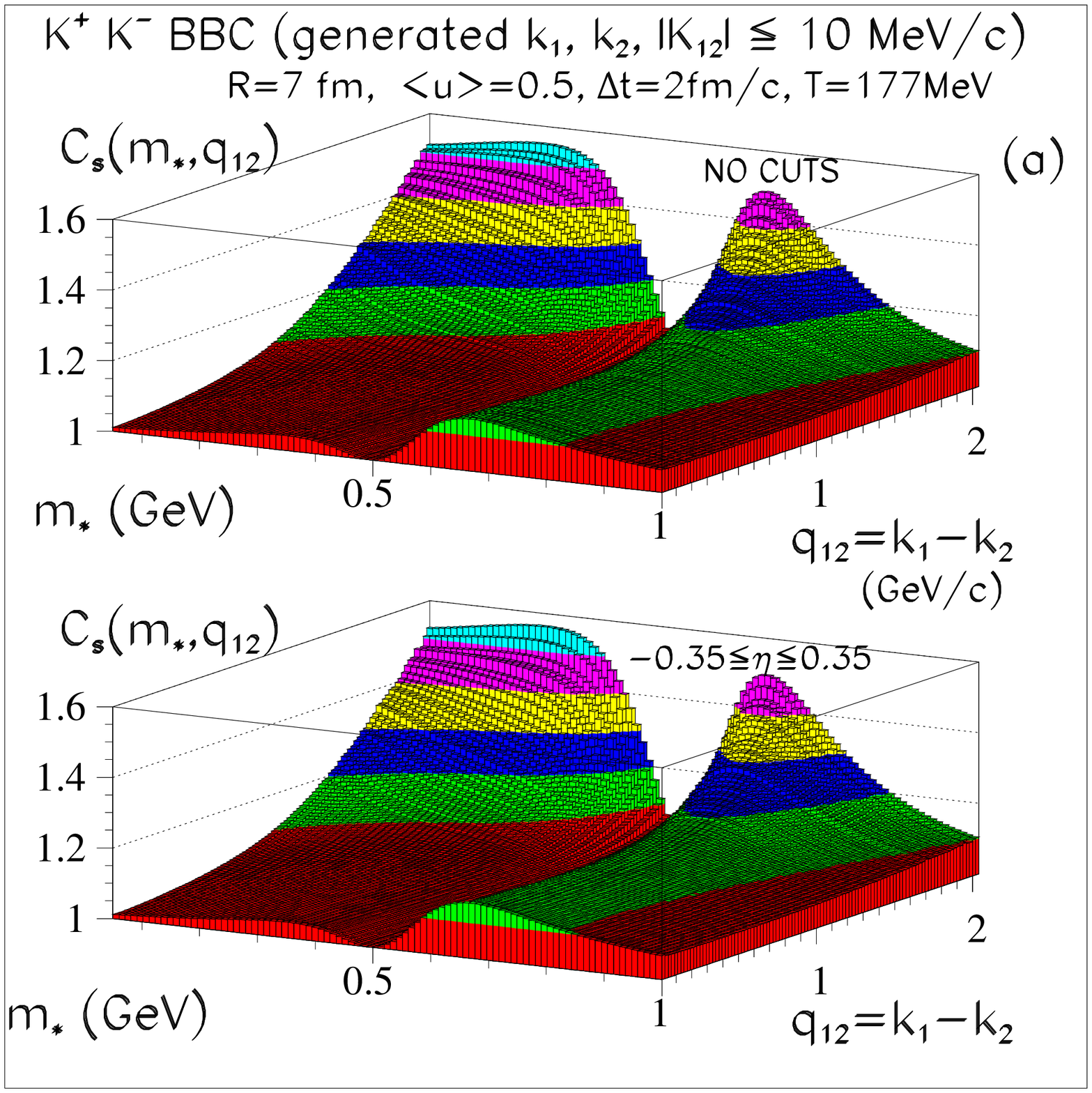}
 \includegraphics[height=.41\textheight]{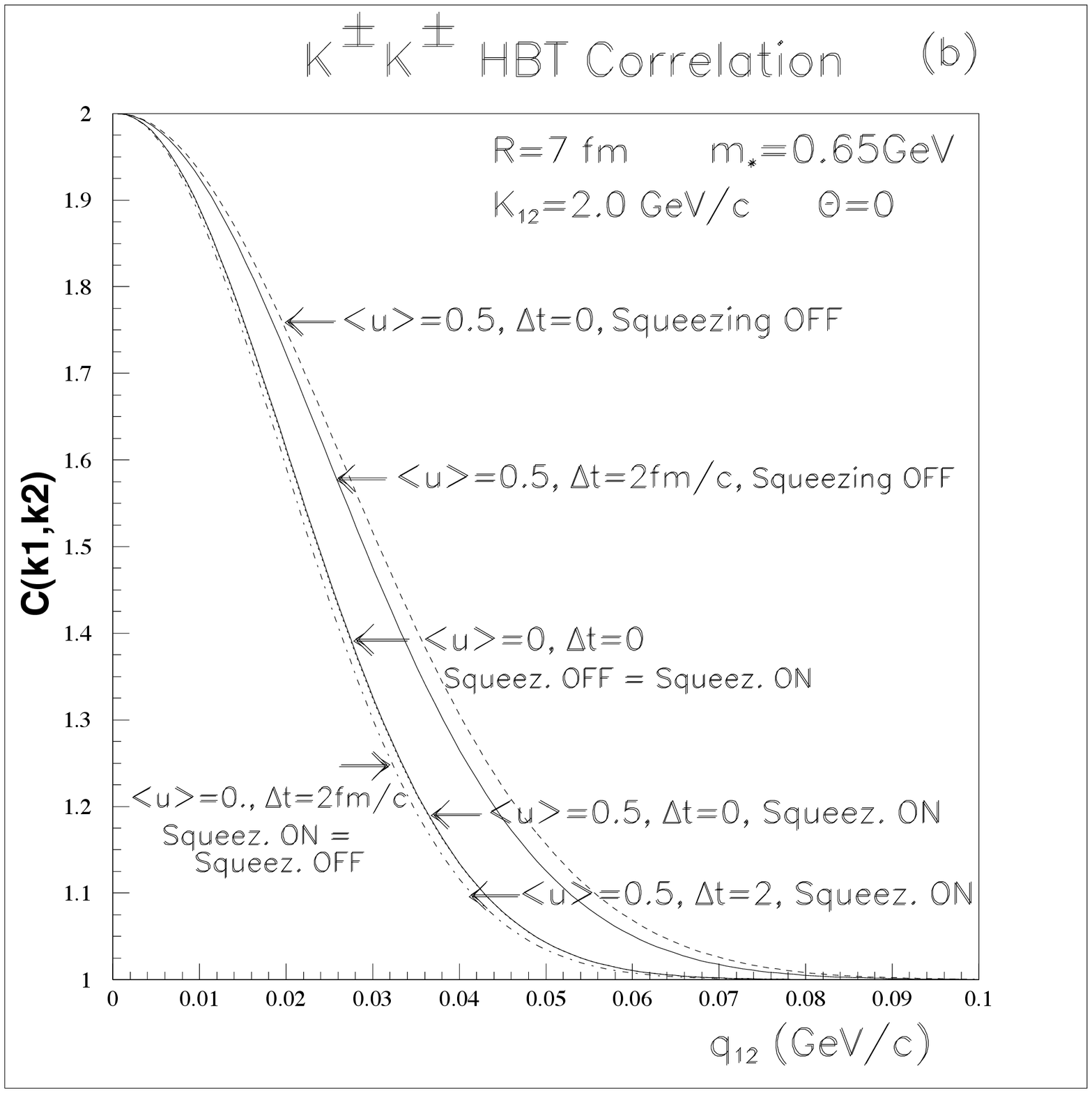}
\caption{Part (a) shows the behavior of the correlation function with the in-medium modified mass and the relative momentum of the pair. Part (b) shows the HBT correlation function when squeezing and radial flow are either absent or present.}
\end{center}
\end{figure}

In the case of the squeezed correlations of $K^+ K^-$ pairs, we  show in Figure 2(a) results for the generated momenta of the pairs within the narrow interval $|\mathbf{K_{_{12}}}| \le 10$ MeV/c, by plotting the squeezed correlation, $C_s(m_*,\mathbf{q_{_{12}}})$ versus $m_*$ and $\mathbf{q_{_{12}}}$. For the kaons, we can fixe the value of the shifted mass to be $m_*\approx 650$ MeV, corresponding to one of the maxima in Figure 2(a), and then procedure similarly to what was done in the $\phi \phi$ case. The result is shown in Figure 3 of Ref.\cite{psdismd08}. 
Also in this case the intensity of the squeezed correlation would be large enough to be searched for experimentally.

Next,  we investigate how the behavior of the identical particle correlations could be affected in case of in-medium mass modification, since the femtoscopic correlation function also depends on the squeezing factor, $f_{i,j}(m,m_*)$. 
The HBT correlation function is obtained by inserting the chaotic amplitude, 
{\small
$G_c(\mathbf{k}_1,\mathbf{k}_2) = \frac{E_{_{1,2}}}{(2\pi)^\frac{3}{2}} \Bigl\{ |s_{_0}|^2 R^3 \exp\Bigl[-\frac{R^2}{2}(\mathbf{k}_1-\mathbf{k}_2)^2\Bigr] + 
n^*_0 R_*^3 (|c_{_0}|^2+|s_{_0}|^2) 
\exp\Bigl[-\frac{(\mathbf{k_1}+\mathbf{k}_2)^2}{8m_* T_*}\Bigr] \exp\Bigl[-\Bigl(\frac{im\langle u\rangle R}{2m_* T_*}\Bigr)(\mathbf{k}_1^2-\mathbf{k}_2^2)\Bigr]  \exp\Bigl[-\Bigl( \frac{1}{8 m_* T} + \frac{R^2_*}{2}\Bigr)(\mathbf{k_1}-\mathbf{k_2})^2\Bigr] \Bigr\}, 
$} 
together with the expression for the spectrum, into Eq.(\ref{fullcorr}). 
We use the case of identical $K^\pm K^\pm$ pairs as illustration, as seen in Figure 2(b). The investigation is extended to both the cases of instant emission ($\Delta t=0$) and finite emission ($\Delta t$=2fm/c). In this figure, we can see the well-known result corresponding to the narrowing of the femtoscopic correlation function with increasing emission times, as well as the broadening the curve with flow in the absence of squeezing, as expected. However, if the squeezing originated in the mass-shift is present, its effects 
tend to oppose to those of flow (for large  $|\mathbf K|$, it practically cancels the broadening of the correlation function due to flow), another striking indication of mass-modification, even in HBT!

\section{Conclusions}

In the present work we suggest an effective way to search for the back-to-back squeezed correlations in heavy ion collisions at RHIC, and later at LHC energies, by investigating the squeezed correlation function, $C_s({\mathbf k_1},{\mathbf k_2})=C_s({\mathbf K},{\mathbf q})$, in terms of $2{\mathbf K_{_{12}}}\!=({\mathbf k_{_{1}}}+{\mathbf k_{_{2}}})$, for different values of ${\mathbf q_{_{12}}}\!=({\mathbf k_{_{1}}}-{\mathbf k_{_{2}}})$.

We showed that, in the presence of flow, the signal is stronger over the momentum regions analyzed in the plots, suggesting that flow may help to effectively discover the BBC signal experimentally. 
Another important point that we find, within this simplified model and in the non-relativistic limit considered here, is that the squeezing would distort significantly the HBT correlation function as well, tending to oppose to the flow effects on those curves, practically neutralizing it for large values of $|\mathbf K|$. 

The analysis in terms of the variable $2\mathbf K$ would not be suited for a genuine relativistic treatment.
In this case, however, a momentum variable could be constructed, as $Q_{back}=(\omega_1-\omega_2,\mathbf k_1 + \mathbf k_2)=(q^0,2\mathbf K)$. In fact, it would be preferable to redefine this variable  as $Q^2_{bbc} = -(Q_{back})^2=4(\omega_1\omega_2-K^\mu K_\mu )$, whose  non-relativistic limit is $Q^2_{bbc} \rightarrow (2 \mathbf K)^2$, as discussed in Ref.\cite{pscn08,psdismd08}. Finally, it is important to emphasize that all the effects and signals discussed here would exist only if the particles analyzed had their masses modified by interactions in the hot and dense medium. 

\bigskip
\noindent
{\bf Acknowledgments}\\
SSP is very grateful to the Organizing Committee of the WPCF 2008 for the kind support to attend the workshop. DMD thanks CAPES and FAPESP for the financial support.

\end{document}